\documentclass[12pt, preprint]{aastex}

\newcommand{\co}{\mbox{$^{12}$CO}}
\newcommand{\coa}{\mbox{$^{13}$CO}}
\newcommand{\cob}{\mbox{C$^{18}$O}}
\newcommand{\kms}{\mbox{km s$^{-1}$}}
\newcommand{\kkms}{\mbox{K km s$^{-1}$}}
\newcommand{\msun}{\mbox{M$_\odot$}}
\newcommand{\vlsr}{\mbox{V$_{LSR}$}}
\newcommand{\etal}{\mbox{et~al.~}}

\shorttitle{FCRAO Survey of the Taurus Molecular Cloud}
\shortauthors{Narayanan et al.}

\begin{document}

\title{The Five College Radio Astronomy Observatory CO Mapping Survey
  of the Taurus Molecular Cloud}
\author{Gopal Narayanan\altaffilmark{1}, 
Mark H. Heyer\altaffilmark{1},
Christopher Brunt\altaffilmark{1,2},
Paul F. Goldsmith\altaffilmark{3},
Ronald Snell\altaffilmark{1},
Di Li\altaffilmark{3}}
\altaffiltext{1}{Dept. of Astronomy, Univ. of Massachusetts, Amherst MA
  01003}
\altaffiltext{2}{University of Exeter}

\altaffiltext{3}{Jet Propulsion Laboratory, California Institute of
  Technology}
\email{gopal@astro.umass.edu}

\begin{abstract}
The FCRAO Survey of the Taurus Molecular Cloud observed the \co\ and
\coa\ J=1-0 emission from 98 deg$^2$ of this important, nearby star
forming region.  This set of data with 45\arcsec\ resolution comprises
the highest spatial dynamic range image of an individual molecular
cloud constructed to date, and provides valuable insights to the
molecular gas distribution, kinematics, and the star formation
process.  In this contribution, we describe the observations,
calibration, data processing, and characteristics of the noise and
line emission of the survey.  The angular distribution of \co\ and
\coa\ emission over 1 \kms\ velocity intervals and the full velocity
extent of the cloud are presented.  These reveal a complex, dynamic
medium of cold, molecular gas.
\end{abstract}

\section{Introduction}
The Taurus Molecular Cloud has long been a valued target for studies
of star formation and the interstellar medium.  Its proximity (140 pc)
and displacement from the Galactic Plane (b$\sim$-19$^\circ$) afford
high spatial resolution views of a star forming region with
little or no confusion from background stars and gas.  However, the
large angular extent of the Taurus cloud ($>$100 deg$^2$) on the sky
has limited most previous investigations to small, targeted areas or
full coverage with coarse angular resolution.  Given the complexity of
the ISM and the star formation process, such limited views may miss
fine, structural features within the cloud or fail to recognize large
scale patterns in the gas.  Yet such structure provides critical clues
to the prevailing physical processes that govern the formation of
stars.

The sheer volume of the data that have been obtained and the number of
analyses that have been carried out preclude giving a complete listing
of the references to Taurus, so we will have to be selective rather
than comprehensive, recognizing that we may have omitted many valuable
contributions.  The most complete inventory of the molecular gas
content within the Taurus cloud is provided by \citet{ungerechts1987},
who observed \co\ J=1-0 emission from 750 deg$^2$ of the
Taurus-Auriga-Perseus regions.  They estimate the molecular mass
resident within the Taurus-Auriga cloud to be 3.5$\times$10$^4$ \msun.
However, the 30\arcmin\ angular resolution of this survey precludes an
examination of the small scale structure of the cloud.  Targeted
studies with higher angular resolution of \coa\ and \cob\ emission
from individual sub-clouds of Taurus reveal some of the relationships
between the molecular gas, magnetic fields, and star formation but
offer little insight to the coupling of these structures to larger
scales and features \citep{schloerb1984, heyer1987, mizuno1995,
  onishi1996}.  Other studies have utilized other molecular tracers
and even higher angular resolution to probe gas having different
characteristics in some limited small regions of Taurus. Some examples
include \cite{langer1995} employing CCS, \cite{onishi1996} and
\cite{onishi1998} using C$^{18}$O, \cite{onishi2002} using
H$^{13}$CO$^+$, and \cite{tatematsu2004} employing N$_2$H$^+$.  Many
individual cores have been observed in NH$_3$, a tracer in which they
appear well-defined, as indicated by the compilation of
\citet{jijina1999}. The Leiden/Dwingeloo 21 cm study
\citep{burton1994} traced the atomic hydrogen towards Taurus, but with
an angular resolution of 35\arcmin.  One investigation
\citep{shuter1987} used the Arecibo radio telescope having an angular
resolution of 4\arcmin, but included only $\sim$1300 positions to
probe the self--absorption seen in the 21 cm HI line.  This cold
atomic hydrogen appears to be associated with molecular gas
\citep{li2003, goldsmith2005}, but the limited sampling of Shuter et
al. does not reveal much about its morphology.  The far--infrared
emission from Taurus has been studied by \cite{abergel1995}, who also
compared it to moderate resolution maps of \coa\ J = 1$\rightarrow$0
emission.  The dust column density distribution has been examined from
2MASS extinction by \cite{padoan2002} and does bear a quite close
resemblance to the integrated intensity of \coa, and thus to the
column density of gas.


With the deployment of heterodyne focal plane arrays at millimeter and
submillimeter wavelengths mounted on large or moderate-sized
telescopes, it is possible to construct high spatial dynamic range,
spectroscopic images of molecular line emission from interstellar
clouds \citep{heyer2000}.  In this contribution, we present the data
from the spectroscopic imaging campaign of \co\ and \coa\ J=1-0
emission obtained with the FCRAO 14m telescope and the 32 element
SEQUOIA focal plane array. The combined dataset of \co\ and
\coa\ allows us to make a much more accurate inventory of gas column
density and mass for the Taurus molecular cloud than hitherto
attempted. At high gas column densities, because of its greater
optical depth, \co\ is insensitive to variations in underlying
column density. However, because of its higher abundance, it is
effective in tracing much of the lower column density regions in the
molecular cloud. The \coa\ transition, with its lower optical depth,
allows us to probe deeper into the envelopes and trace the underlying
structure of the molecular gas. Using both \co\ and \coa\ we can
derive the structure and column density of the molecular gas in
different regimes of column density. The data from this survey
constitute the most detailed view constructed to date of a molecular
cloud and provides critical information for detailed studies of gas
dynamics within the molecular interstellar medium. In this paper, we
summarize the instrumentation, and data collection and analysis
methods, and present global characteristics of noise and line emission
from this survey. Detailed scientific motivation and results can be
found in \citet{goldsmith2008}.

\section{Observations}
\subsection{Instrumentation}

All observations presented in this paper were taken with the 14~meter
diameter millimeter-wavelength telescope and the 32 pixel focal plane
array SEQUOIA \citep{erickson1999} of the Five College Radio Astronomy
Observatory.  The FWHM beam sizes of the telescope at the observed
frequencies are 45\arcsec\ (115.271202 GHz) and 47\arcsec\ (110.201353
GHz). The main beam efficiencies at these frequencies are 0.45 and
0.50 respectively as determined from measurements of Jupiter. Previous
measurements of the extended error beam of the telescope and radome
structure by measuring the disks of the sun and moon indicate that
there can be $\sim 25$\% net contribution from extended emission
outside the main beam from a region $\sim 0.5\arcdeg$ in diameter. The
shape of this error beam is approximately circular, but the amount of
contribution of emission at any given point from this error beam
pattern depends on details of the distribution of the emission from
the source. All data presented here are in T$_A^*$ (K), uncorrected
for telescope beam efficiencies. The backends were comprised of a
system of 64 autocorrelation spectrometers each configured with 25 MHz
bandwidth and 1024 spectral channels.  No smoothing was applied to the
autocorrelation function so the spectral resolution was 29.5 kHz per
channel corresponding to 0.076 \kms\ (\co) and 0.080 \kms (\coa) at
the observed frequencies. The total coverage in velocity is 65
\kms\ (\co) and 68 \kms\ (\coa) respectively. The spectrometers were
centered at a v$_{\rm LSR}$ of 6 \kms.


\subsection{Data Collection}

The Taurus Molecular Cloud was observed over two observing seasons
starting in November 2003 and ending in May 2005.  The \co\ and
\coa\ lines were observed simultaneously enabling excellent positional
registration and calibration.  System temperatures ranged from 350-500
K for the \co\ line and 150-300 K for the \coa\ line.  The fiducial
center position of the map was $\alpha$(2000) = $04^h32^m44^s.6$,
$\delta$(2000) = $24\arcdeg 25\arcmin 13\arcsec.08$.  The $\sim 98$
deg$^2$ selected region of the cloud was divided into 356 submaps,
each $30\arcmin\times 30\arcmin$ in size. Each submap was observed
using the On-The-Fly (OTF) mapping technique in which the telescope is
continuously scanned back and forth along the Right Ascension axis
while rapidly reading the spectrometers.  The telescope was scanned at
a rate such that 2 samples of data were collected while traversing the
FWHM beam width at 115 GHz (45\arcsec). An additional ramping offset
of 1\arcmin\ at the beginning and end of each scan was used to ensure
stable motion of the antenna.  OTF scans were spaced by 34\arcsec\ in
declination.  Since the SEQUOIA dewar is fixed with respect to the
Azimuth-Elevation coordinate system, the scanning rate and declination
offset between scan rows assured that the target field was densely
sampled by the array.  A given position on the sky was observed by
most of the pixels in the array.  Such redundant measurements increase
the effective integration time at that position and also dilute the
effects that arise from the small variations of noise and gain between
the pixels of the array.

With a focal plane array and ramping offsets, there are data points
that lie outside of the target 30\arcmin$\times$30\arcmin\ area.  Upon
convolution into a regularly spaced grid (see \S\ref{data_process}),
these peripheral points contribute to the integration times of points
located within but near the edge of the target area.  To increase our
mapping efficiency the submaps were spaced by 33\arcmin, which enabled
the peripheral spectra from contiguous submaps to be coadded with
sufficient integration time to achieve the desired sensitivity.  The
target T$_A^*$ sensitivities, $\sigma$, for each convolved (see \S2.3)
\co\ and \coa\ spectrum were 0.65 K and 0.25 K respectively as gauged
by the root-mean square of antenna temperature values within
signal-free backend channels.  The sensitivity of a submap was
evaluated by the median rms value of $\sigma$, from the set of
constituent spectra.  Submaps with rms values in excess of 0.75 K were
re-observed until the median value of $\sigma$ of the co-added set of
data was less than this maximum allowed value.

Position-switched spectral data requires a clean OFF position, which
is free of line emission over the entire footprint of the SEQUOIA
array. We used 5 different OFF positions spread around the Taurus
molecular cloud. The right ascension and declinations of the OFF
positions used in J2000 co-ordinates are ($04^h20^m00^s$, $23\arcdeg
00\arcmin 00\arcsec$), ($04^h42^m00^s.9$, $23\arcdeg 25\arcmin
41\arcsec$), ($04^h54^m22^s.7$, $24\arcdeg 43\arcmin 28\arcsec$),
($04^h01^m15^s$, $27\arcdeg 32\arcmin 00\arcsec$), and
($04^h20^m05^s.01$, $32\arcdeg 24\arcmin 59\arcsec.2$). For any given
submap observed, we typically chose the closest OFF position. The
first two positions were known to be free of CO emission from previous
studies. The other positions were derived from constructing small maps
in \co\ and \coa\ at each position (a guess at an initial OFF was
derived from the IRAS 100$\mu$m map). These maps were typically
10$\arcmin\times$ 10$\arcmin$ in size, and it was verified that
\co\ was not detectable to $\sim 0.25$ K level in the map.

In addition to the OTF maps, for each day of observing, pointing and
focus measurements were made using a SiO maser source, IKTau
($03^h53^m28^s.81$, $11\arcdeg 24\arcmin 22\arcsec.60$) every
few hours. The pointing accuracy of the FCRAO 14~m telescope is
typically better than 5\arcsec. After each pointing observation, a
single 80 second long position-switched observation in \co\ and
\coa\ was performed at the fiducial center position to keep track of
the relative calibration accuracy of the whole survey (see
\S\ref{ps-uncertainty}).

For the observing campaign, a dynamic website with a relational
database backend was created. From this website, the observing team
could monitor the status of observations made to date, plan new
observations, generate observing scripts, immediately reduce the OTF
map after it was completed using a simple data reduction pipeline,
view the integrated intensities of the spectra at the fiducial center
position, and produce summary statistics of the map. The website also
listed the sensitivities obtained in each submap, and produced
prioritized list of submaps left to observe.

\subsection{Data Processing}
\label{data_process}
On-the-Fly mapping generates a set of data that is densely but
irregularly sampled on the sky.  To construct regularly sampled
spectroscopic data cubes and to coadd spatially redundant
measurements, the data were convolved into an output grid using a
kernel that accounts for the edge taper of the 14m antenna to minimize
noise aliasing, and retains the full angular resolution of the
telescope.  In addition to this spatial weighting, a spectrum was
further weighted by the factor $\sigma^{-2}$, where $\sigma$ is the
measured rms noise. The spatial pixel separation in the regridded
maps is $20\arcsec$.

Following this initial processing of submaps into data cubes, a
systematic trend was present in the spectral baselines owing to gain
drifts in the receiver.  Each OTF scan begins with the measurement of
OFF position, followed by two raster scanned rows of ON position
observations, finally ending in an OFF position observation. The final
spectra were computed using both OFF position observations weighting
them equally. Spectra taken close in time to either of the OFF
positions typically show a concave curvature in the baseline. While
spectra taken approximately midway between the two OFFs typically show
a convex curvature in baseline. These curvatures were small with
respect to the noise of any single spectrum but were evident when
averaging spectra with comparable time displacements from the off
measurement or when integrating individual spectra over velocity
intervals comparable to the half period of the baseline curvature.

To rectify these baseline curvatures, we subtracted a baseline
polynomial that depended on the spectral and sky coordinates of
spectra within a given submap.  We found that a second order
polynomial along the spectral axis was sufficient and variations along
the spatial axes could be adequately described by a first order
polynomial.  Accordingly, for each submap, we fit the parameterized
baseline polynomial,
$$T_{base}(x,y,v)=\sum_{i=0}^2 (v-v_\circ)^i f_i(x-x_\circ)
g_i(y-y_\circ) \eqno(1) $$ where $x_\circ,y_\circ,v_\circ$ are the
central coordinates of the submap, and $f_i=a_i+b_i(x-x_\circ)$ and
$g_i = c_i+d_i(y-y_\circ)$ parameterize the variation of the baseline
curvature over the spatial axes.  In the fitting procedure, we exclude
the window $-5 < V_{LSR} < 15$ \kms\ that contains the spectral line
emission from the Taurus Molecular Cloud.

This procedure has the advantage that only 12 parameters are required
to specify the baselines for an entire submap, rather than the three
per spectrum that would be needed for a standard baseline subtraction.
All of the parameters are recorded so that the unbaselined data could
be recovered if necessary.  The fitting of a smooth baseline function
to an entire submap also avoids the inadvertent removal of a broad,
low-level, localized emission line component that could be produced by
outflows and other energetic processes.

\begin{figure*}[h]
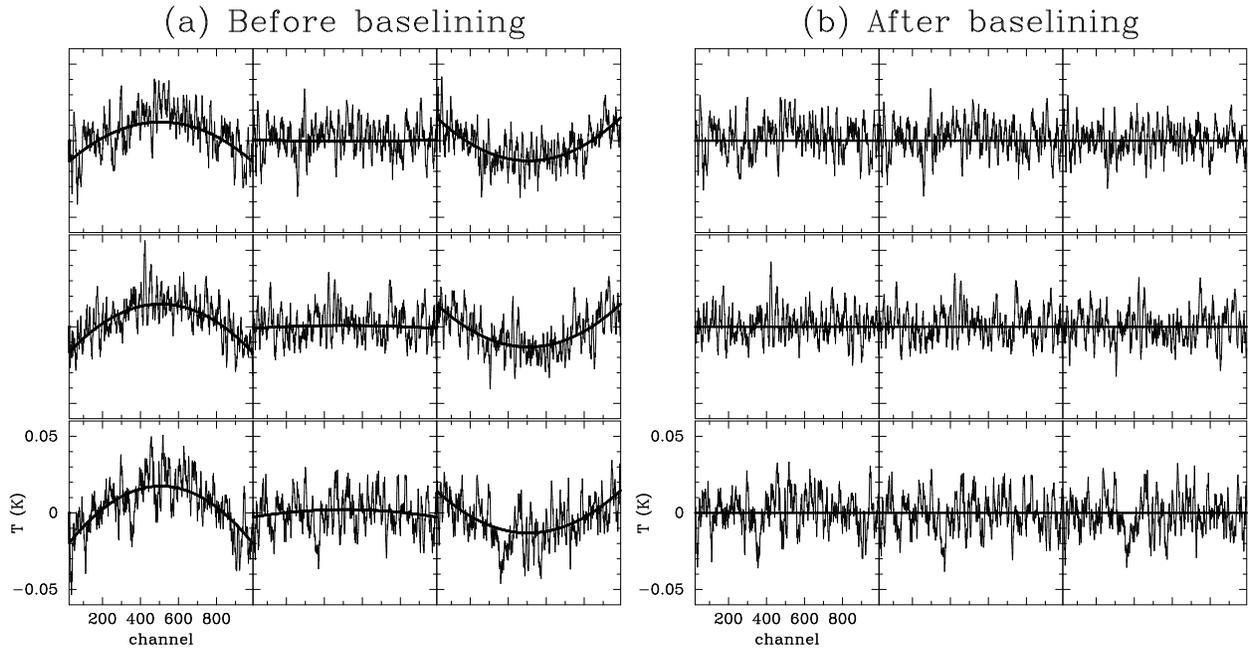

\begin{center}
\includegraphics[angle=-90,scale=0.5]{f1a.eps}
\includegraphics[angle=-90,scale=0.5]{f1b.eps}
\figcaption{(a) $^{13}$CO spectra in a $30\arcmin\times 30\arcmin$
  region averaged over $10\arcmin\times 10\arcmin$ and fitted
  parameterized baselines (thick lines). The rightmost column of
  spectra corresponds to observations taken close in time to an OFF
  measurement, while the leftmost column corresponds to spectra taken
  farthest in time from an OFF observation. The middle column
  represents a time in between these extremes. See the text for
  details. (b) The same box--averaged spectra after baseline removal,
  with thick lines at zero intensity for reference.\label{baseline}}
\end{center}
\end{figure*}

The maximum amplitudes of the baseline curvature were typically greater
in the \coa\ data, where the most severe cases were still limited to
levels less than 25 mK ($\sigma/10$).  To demonstrate the need and
effect of this baselining procedure, we show in Figure~\ref{baseline}a
\coa\ spectra averaged over 10\arcmin$\times$10\arcmin\ boxes from a
submap representative of the worse cases of baseline curvature.  We
emphasize that the curvature is not readily evident in a single
spectrum.  The thick solid lines show the fitted parameterized
baselines averaged over the same boxes. The spatial variation in
curvature is most significant along the scanning ($x$) direction, as
noted above, with a slow drift in the $y$ direction as the ON-OFF
elevation difference varies during the 2-hour submap
observation. Figure~\ref{baseline}b shows the same average spectra
after removal of the fitted baselines, with thick lines at zero
intensity for reference.

The 356 regridded submaps, each of which have a size of $\sim
30\arcmin\times 30\arcmin$, were further consolidated into a set of
larger data cubes. There are 88 such consolidated cubes, forming a
grid of $11\times 8$ data cubes, dubbed hard--edge cubes (as they do
not have overlapping regions between two contiguous cubes). Each
'hard-edge' cube is assembled from a set of input regridded
$30\arcmin\times 30\arcmin$ cubes, after removing the spectral and
spatially derived baselines described above from each cube, and
subsequently averaging the data together, weighting them by
$\sigma^{-2}$, where $\sigma$ is the rms in the derived baseline. In
angular offsets, the full extent of the combined hard--edge cubes are
(5.75\arcdeg, $-5.75$\arcdeg) in RA offsets and ($-2.75$\arcdeg,
5.75\arcdeg) in Dec offsets from the fiducial center of the map. Thus
for the full $11.5\arcdeg\times 8.5\arcdeg$ region spaced at
20\arcsec, there are 3,167,100 spectra in each isotopologue in the
combined set of hard--edge cubes. Most of the hard--edge cubes have a
spatial size of 1 square degree, except for the cubes that lie on the
four edges of the region covered, which measure 1.25 square
degrees. The hard--edge cubes at the four corners of the Taurus map
have a size of 1.5625 square degrees ($1.25\arcdeg\times
1.25\arcdeg$).

\section{Results}
\subsection{Calibration Uncertainty in Data}
\label{ps-uncertainty}

\begin{figure}
\begin{center}
\plotone{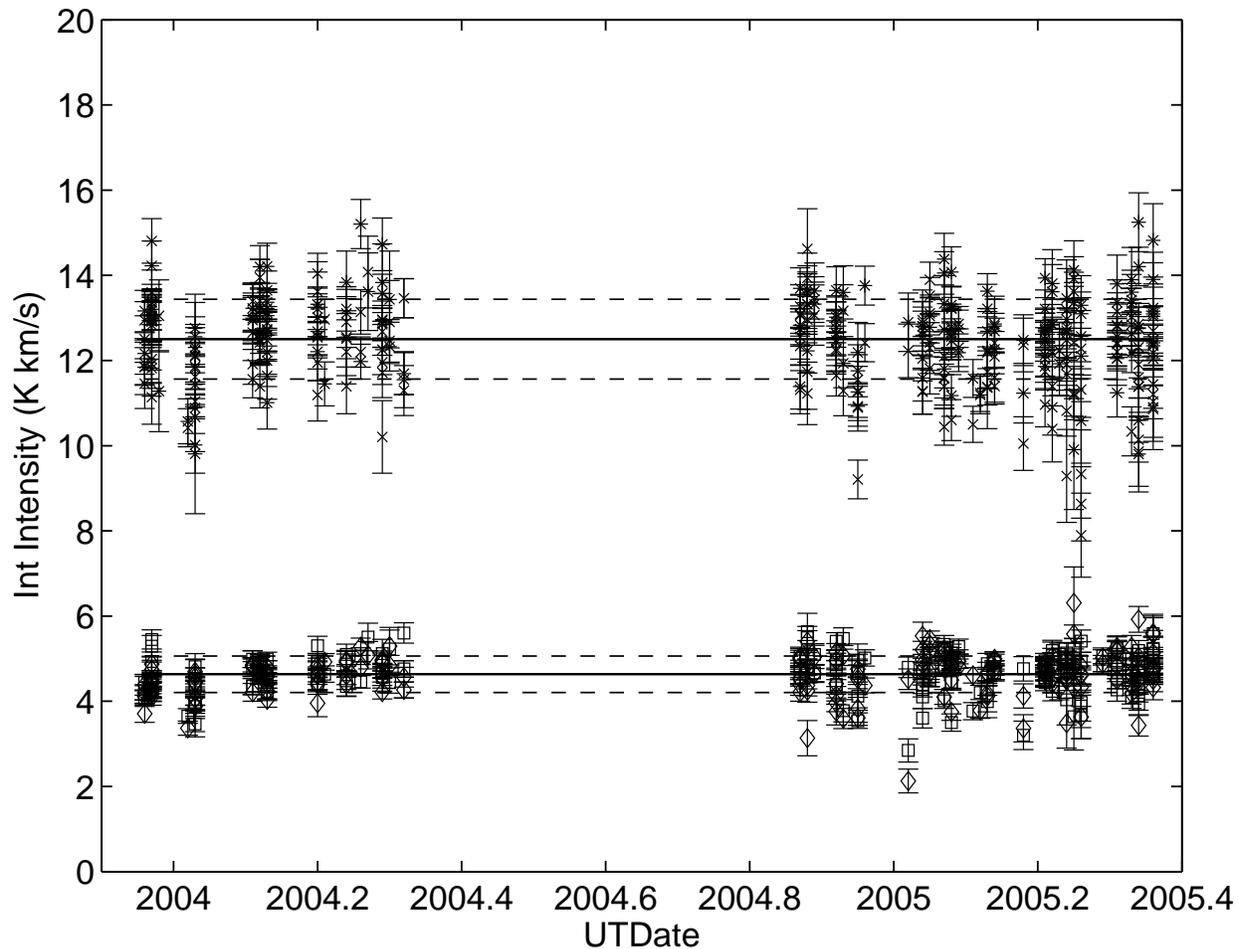}
\figcaption{Integrated intensity in the position-switched spectra for the
  (0,0) position for both polarizations and both isotopologues over
  the entire observing period.  The $^{12}$CO data have integrated
  intensity $\simeq$ 12.5 \kkms and the $^{13}$CO data have integrated
  intensity $\simeq$ 4.6 \kkms. The error bars denote $1\sigma$
  statistical errors in these values.  The $^{12}$CO data are denoted
  by crosses and plus symbols and the $^{13}$CO data by squares and
  diamonds, for the two polarizations. The horizontal solid lines show the mean
  integrated intensities derived for each data set and the dashed
  lines indicate $\pm$ standard
  deviation about these mean values.  The gap in coverage in the
  middle of year 2004 is due to the normal summer shutdown period of
  the FCRAO 14~m telescope. \label{pscalib}}
\end{center}
\end{figure}

In order to track the relative calibration of the survey that was
taken over the course of two observing seasons, we made
position-switched (PS) measurements on the central (0,0) position
every few hours.  Since SEQUOIA does not rotate with the sky, only the
(0,0) position was uniformly repeated.  These (0,0) spectra were
baseline subtracted and integrated over the velocity intervals 2-10
\kms\ and 3-9 \kms\ for \co\ and \coa\ respectively to produce an
integrated line intensity. Both the line profile shape and integrated
intensity offer a cross check on the pointing and focus state of the
telescope system.  The set of these measurements also provide a
quantitative measure of the calibration uncertainty of the
data. Figure~\ref{pscalib} shows the full set of measurements taken
over the course of the survey.  The mean and standard deviation,
weighted by the statistical error for each spectrum, of the integrated
intensity for both isotopologues are $12.5 \pm 0.93$ \kkms\ and $4.6
\pm 0.43$ \kkms. After subtracting the statistical uncertainties in
quadrature, the estimated calibration uncertainties for \co\ and
\coa\ are $0.87$ and $0.43$ \kkms\ respectively, corresponding to a
relative calibration uncertainty of 7\% for \co\ and 9.3\% for \coa.
Submaps that followed or preceded position switched measurements that
significantly deviated from these mean values were reobserved.

\section{Noise Properties}
Quantitative measures of cloud structure rely on the
sensitivity of the data, its uniformity across the target field, and
the underlying noise characteristics of the data.  For individual
submaps, On-the-Fly Mapping with a focal plane array produces near
uniform sensitivity over most of the field.  Toward the edges of the
submap, the noise necessarily increases as there is less integration
time accumulated at these positions.  However, our placement of
submaps ensured that there are sufficient overlaps of contiguous
fields so that these spectra accumulated additional integration time.
Indeed, upon the construction of the 1 deg$^2$ cubes, the sensitivity
of these ``edge'' spectra is comparable or better than the spectra
from the central parts of any submap.

\begin{figure}
\begin{center}
\plotone{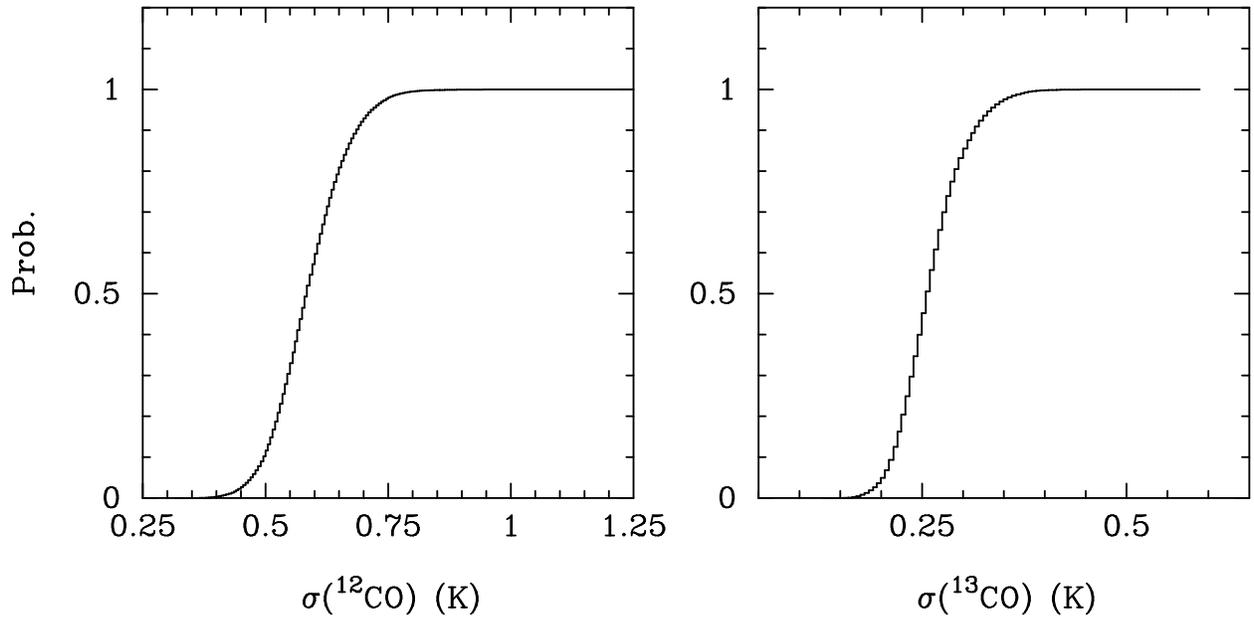}
\figcaption{Cumulative distributions of rms values, $\sigma(^{12}$CO) and 
$\sigma(^{13}$CO) derived from signal--free channels within the 
Taurus \co\ and \coa\ data cubes.
\label{hist_rms}}
\end{center}
\end{figure}

For a given spectrum, the statistical error of the antenna temperature
at any channel, $\sigma$, is conventionally estimated from the
standard deviation of antenna temperatures within intervals in which
no signal is present.  This measure includes noise contributions from
the instrument and atmosphere.  For the Taurus Survey, we have
calculated $\sigma$ for all the co-added convolved and resampled
\co\ and \coa\ spectra excluding values within the velocity range of
the Taurus cloud (-5 $< V_{LSR} < +15$ \kms).  The cumulative
distributions of $\sigma(^{12}CO)$ and $\sigma(^{13}CO$) are shown in
Figure~\ref{hist_rms}.  The steepness of these distributions provides
an approximate measure of noise uniformity.  The first, second, and
third quartile values are 0.53 K, 0.58 K, and 0.63 K for \co\ and 0.23
K, 0.26 K, and 0.28 K for \coa.

\begin{figure}
\begin{center}
\plotone{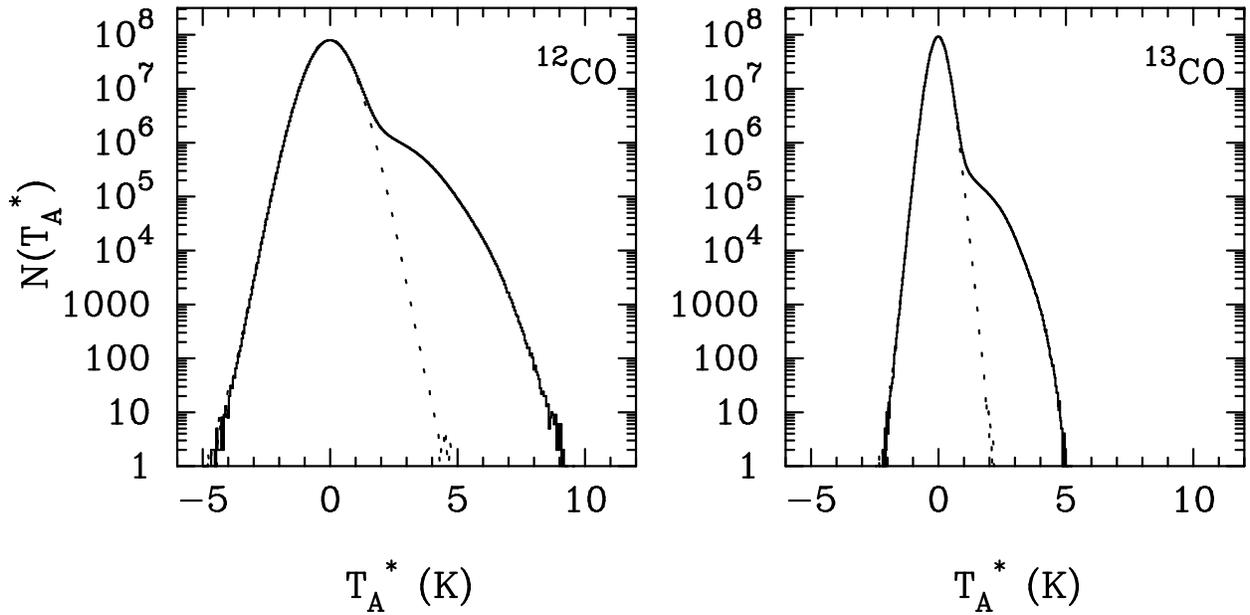}
\figcaption{The distribution of voxel values within 
the composite \co\ and \coa\ data cubes (solid lines).  The dotted lines
show the distribution of values obtained from data cubes of equal size
containing only gaussian noise with values of $\sigma$ that follow the same 
distribution shown in Figure~\ref{hist_rms}.  These provide an excellent 
fit to the data and demonstrate the near gaussian character of 
the noise of individual spectra. 
\label{taurus_voxel}} 
\end{center}
\end{figure}

The antenna temperature distribution of all voxels within the data
cube offers another measure of the noise properties.  In
Figure~\ref{taurus_voxel}, this distribution is shown for the
composite \co\ and \coa\ data cubes of the survey.  For both data
cubes, statistical noise contributes to the peak component centered on
T$_A^*$=0.0 K.  Signal from the Taurus cloud is responsible for the
excess positive emission.  In the ideal case in which all spectra have
gaussian distributed fluctuations with constant rms value, $\sigma$,
the dispersion of the voxel distribution about T$_A^*$=0 K would be
equivalent to $\sigma$.  More realistically, the noise is not uniform
and not necessarily gaussian.  For example, the use of a common
reference position for many source spectra introduces correlated
noise. With OTF mapping using a focal plane array, where a given
position on the sky is sampled by many pixels in the array, this
effect is present but not as severe as discrete mapping methods that
share reference positions \citep{heyer1998, jackson2006}.  As shown in
Figure~\ref{hist_rms}, the noise of this survey is not uniform.  To
assess departures from the gaussian character of the distribution of
noise in the \co\ and \coa\ data, we have generated equivalently sized
data cubes filled with gaussian noise values that follow the same
distribution of $\sigma$ values shown in Figure~\ref{hist_rms}.  The
respective distributions of these purely gaussian noise spectra, shown
as the dotted lines in Figure~\ref{taurus_voxel}, provide an excellent
match to the zero-centered noise components.  We conclude that the
noise measured in the \co\ and \coa\ data cubes can not be readily
distinguished from pure gaussian noise.  It is dominated by
statistical fluctuations from the receiver and the sky rather than by
systematic contributions.

\section{Images}

Two dimensional images are generated from the spectroscopic data cubes
by reducing the information of each spectrum to a single scalar value.
These reductions include maps of maximum value, the integration of
spectra over a set of velocity intervals (zero moment), and higher
moments such as the centroid velocity or line width.  To convey a
fraction of the information resident within these data cubes, we show
some of these reductions in images of the Taurus Cloud.

\begin{figure}
\begin{center}
\epsscale{0.7}
\plotone{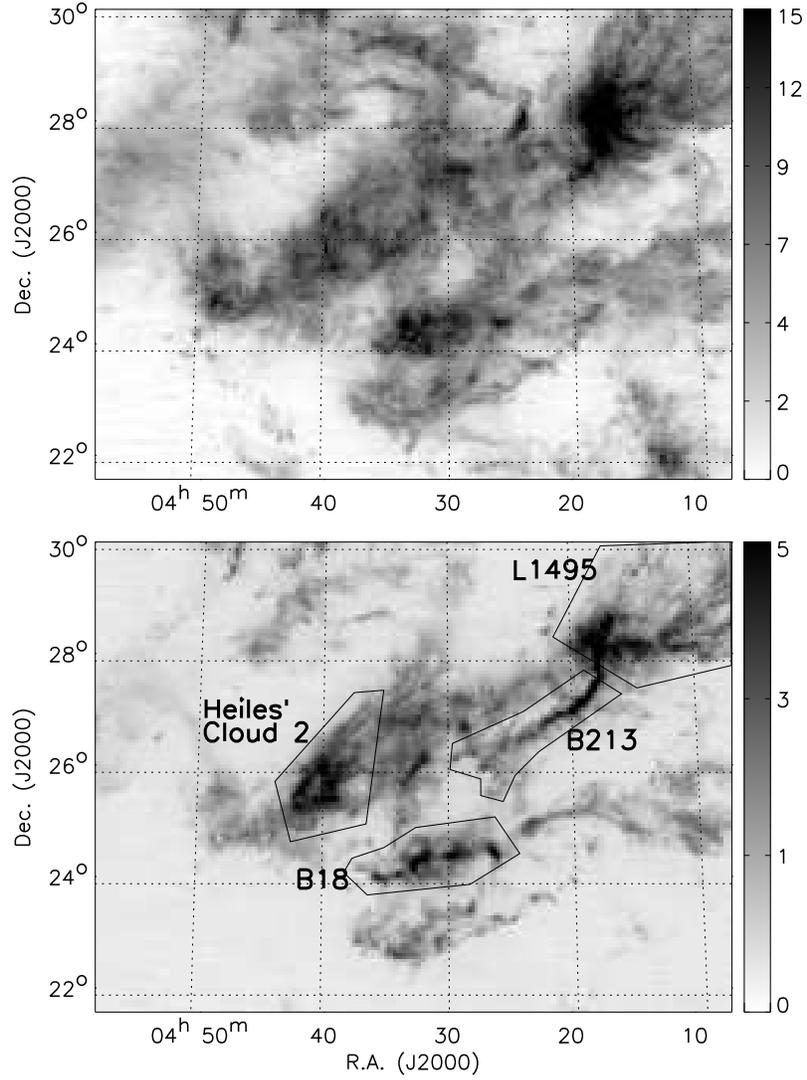}
\figcaption{Integrated Intensity Images for $^{12}$CO (top) and $^{13}$CO
  (bottom). The images are obtained over $-5$ to $20$ \kms\ and $3$ to
  $9$ \kms\ for \co\ and \coa\ respectively. The colorbar on the right
  shows the integrated intensity scale in K.\kms. In the
  \coa\ figure, we also overlay outlines of a few well-known regions
  in Taurus as designated by \citet{onishi1996}.\label{tint}}
\end{center}
\end{figure}

\begin{figure}
\begin{center}
\epsscale{0.7}
\plotone{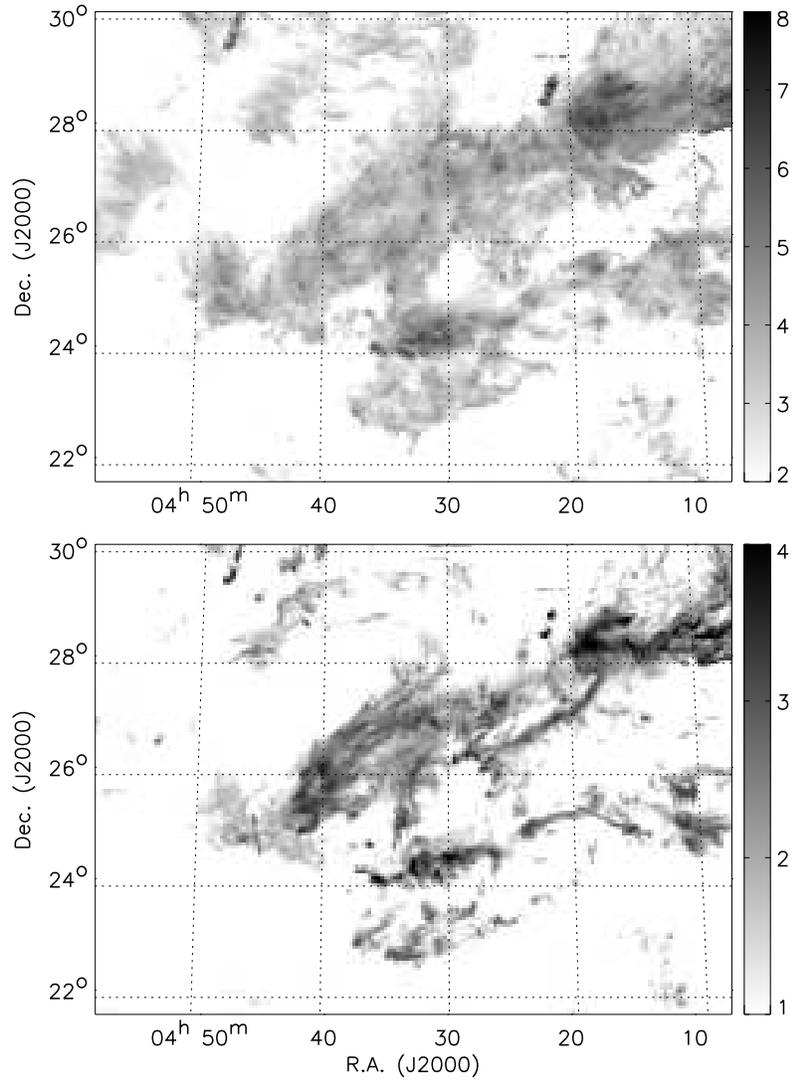}
\figcaption{Images of peak intensity, T$_{max}$, for $^{12}$CO (top) and
  $^{13}$CO (bottom). The peak temperature in the velocity range of 0
  to 12 \kms\ is used in both cases. The colorbar on the right shows
  the T$_{max}$ scale in antenna temperature units
  (K). \label{tmax}}
\end{center}
\end{figure}

Zero moment images derived from the \co\ and \coa\ data cubes over the
\vlsr\ range $0$ to $12$ \kms\ in 1 \kms\ intervals are presented in
Figure~\ref{tint}.  The two images show quite distinct distributions
between the two isotopologues.  The \co\ J=1-0 integrated emission is
mostly diffuse, even towards the three known primary sub clouds of
Taurus (Heiles' Cloud 2, Barnard 18, L1495).  In the most faint
regions, there are weak ``streaks'' or striations of \co\ emission
that are typically aligned with the local magnetic field direction
\citep{goldsmith2008, heyer2008}.  The \coa\ J=1-0 emission is mostly
distributed within high contrast filaments.  However, even within the
diffuse regions, the weak \coa\ emission exhibits a striated pattern.
These differences between the \co\ and \coa\ distributions can be
readily attributed to the higher opacity of the \co\ emission that
likely limits its probed volume to the low column density envelope of
the cloud.  Images of peak \co\ and \coa\ intensity are shown in
Figure~\ref{tmax}.  For the optically thick \co\ line, the peak
intensity is a valuable measure of the excitation temperature.

\begin{figure}
\begin{center}
\epsscale{0.7}
\plotone{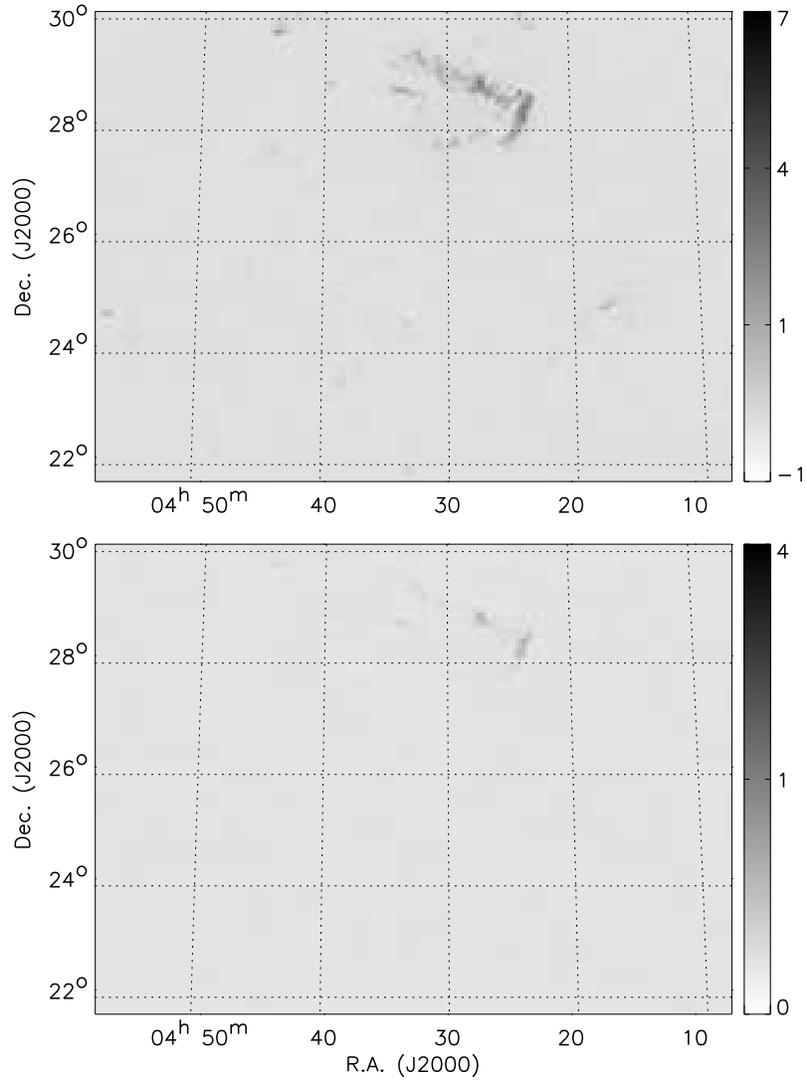}
\figcaption{Images of $^{12}$CO (top) and
  $^{13}$CO (bottom) emission integrated between \vlsr\ 0 to 1 \kms. \label{taurus_00}}
\end{center}
\end{figure}

\begin{figure}
\begin{center}
\epsscale{0.7}
\plotone{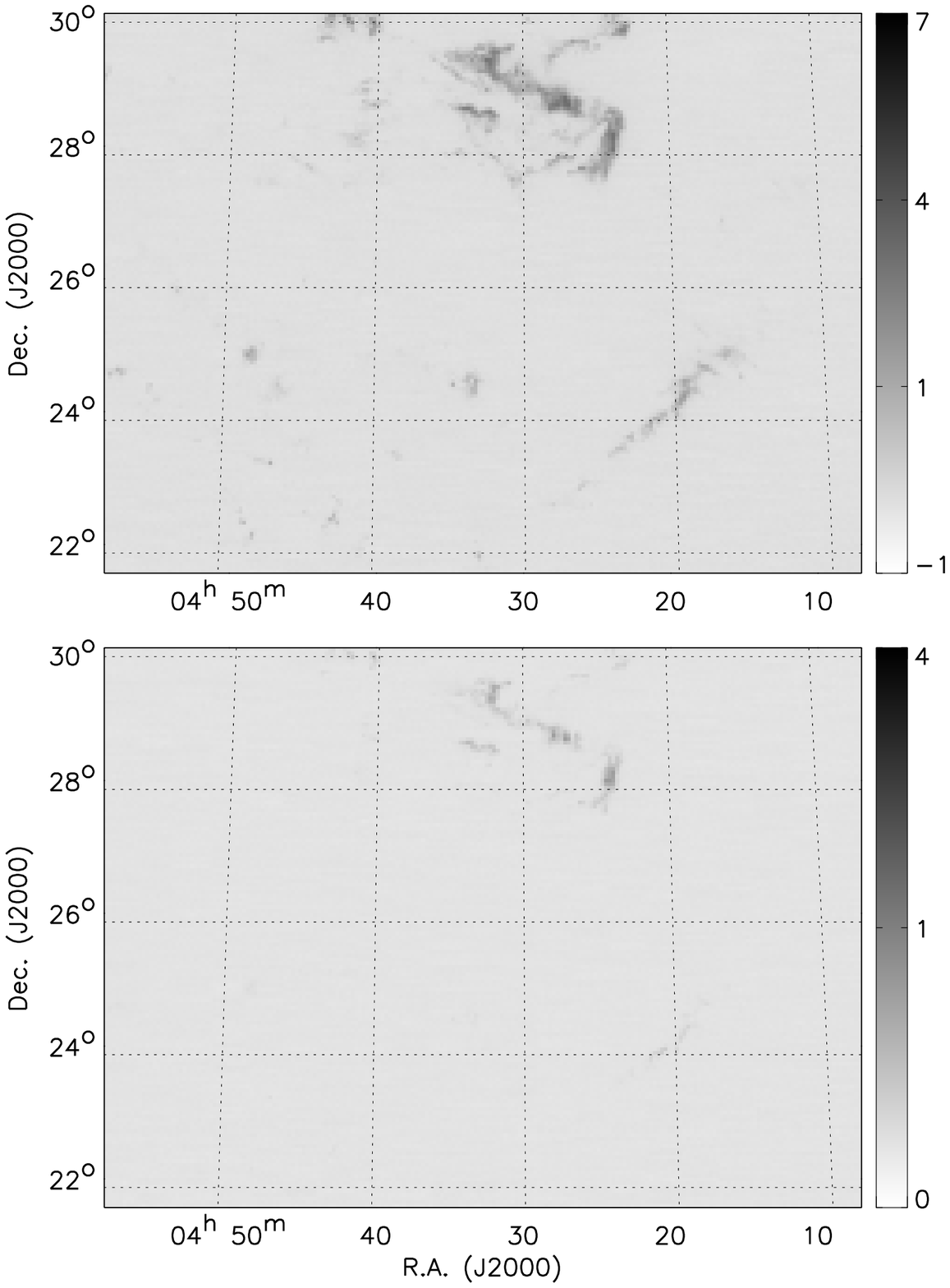}
\figcaption{Same as Figure~\ref{taurus_00} for \vlsr\ 1 to 2 \kms.\label{taurus_01}}
\end{center}
\end{figure}

\begin{figure}
\begin{center}
\epsscale{0.7}
\plotone{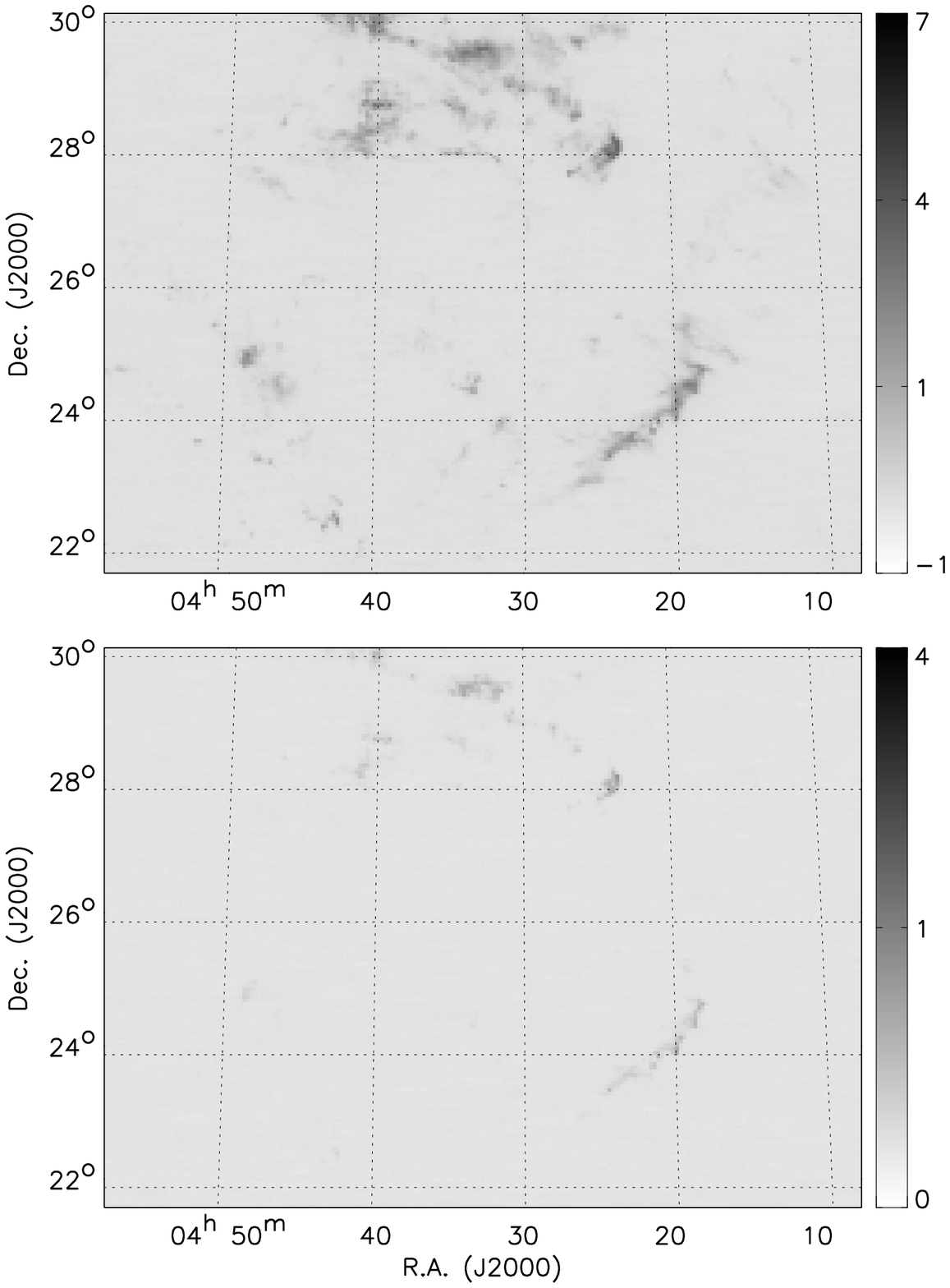}
\figcaption{Same as Figure~\ref{taurus_00} for \vlsr\ 2 to 3 \kms. \label{taurus_02}}
\end{center}
\end{figure}

\begin{figure}
\begin{center}
\epsscale{0.7}
\plotone{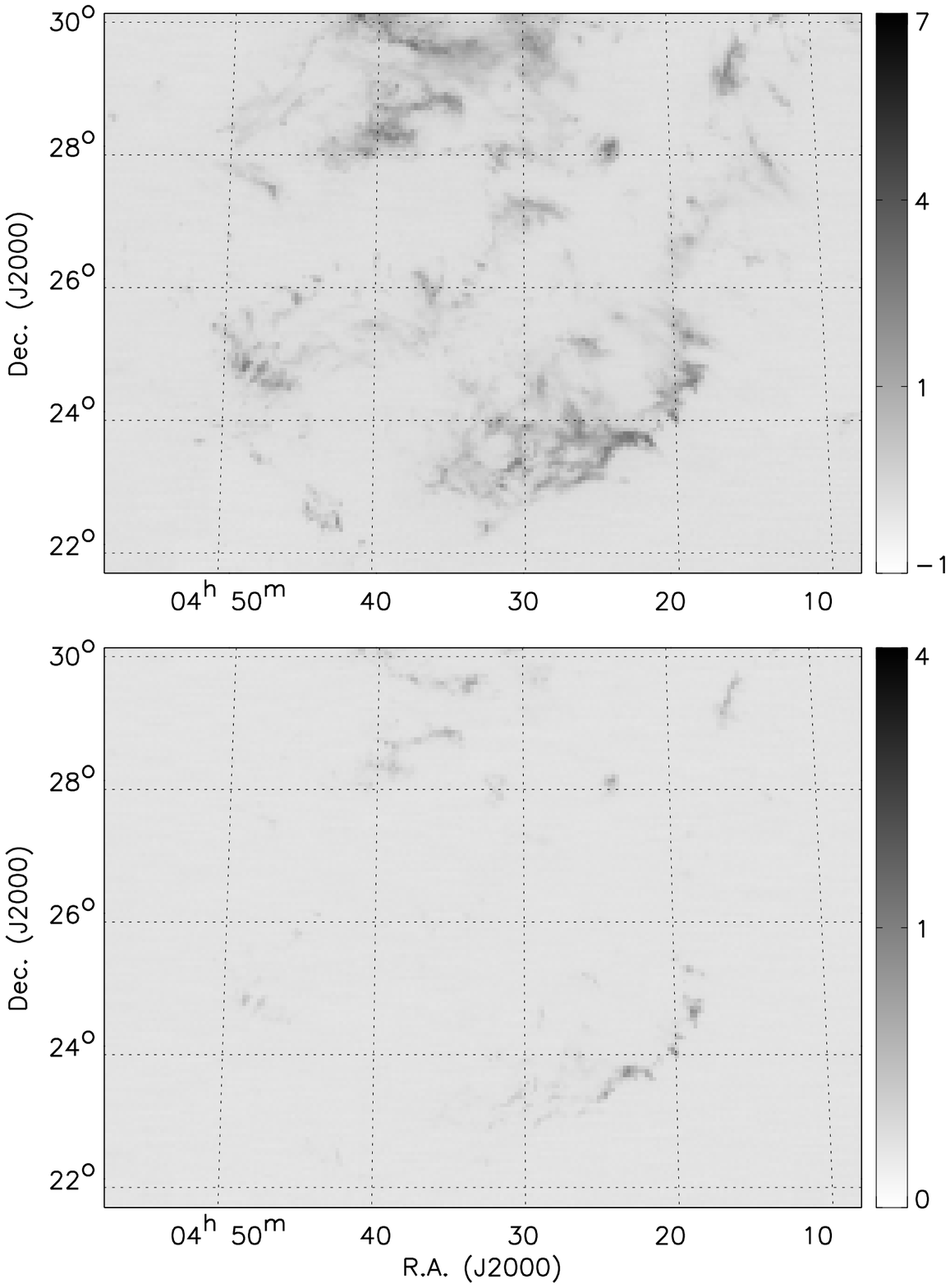}
\figcaption{Same as Figure~\ref{taurus_00} for \vlsr\ 3 to 4 \kms. \label{taurus_03}}
\end{center}
\end{figure}

\begin{figure}
\begin{center}
\epsscale{0.7}
\plotone{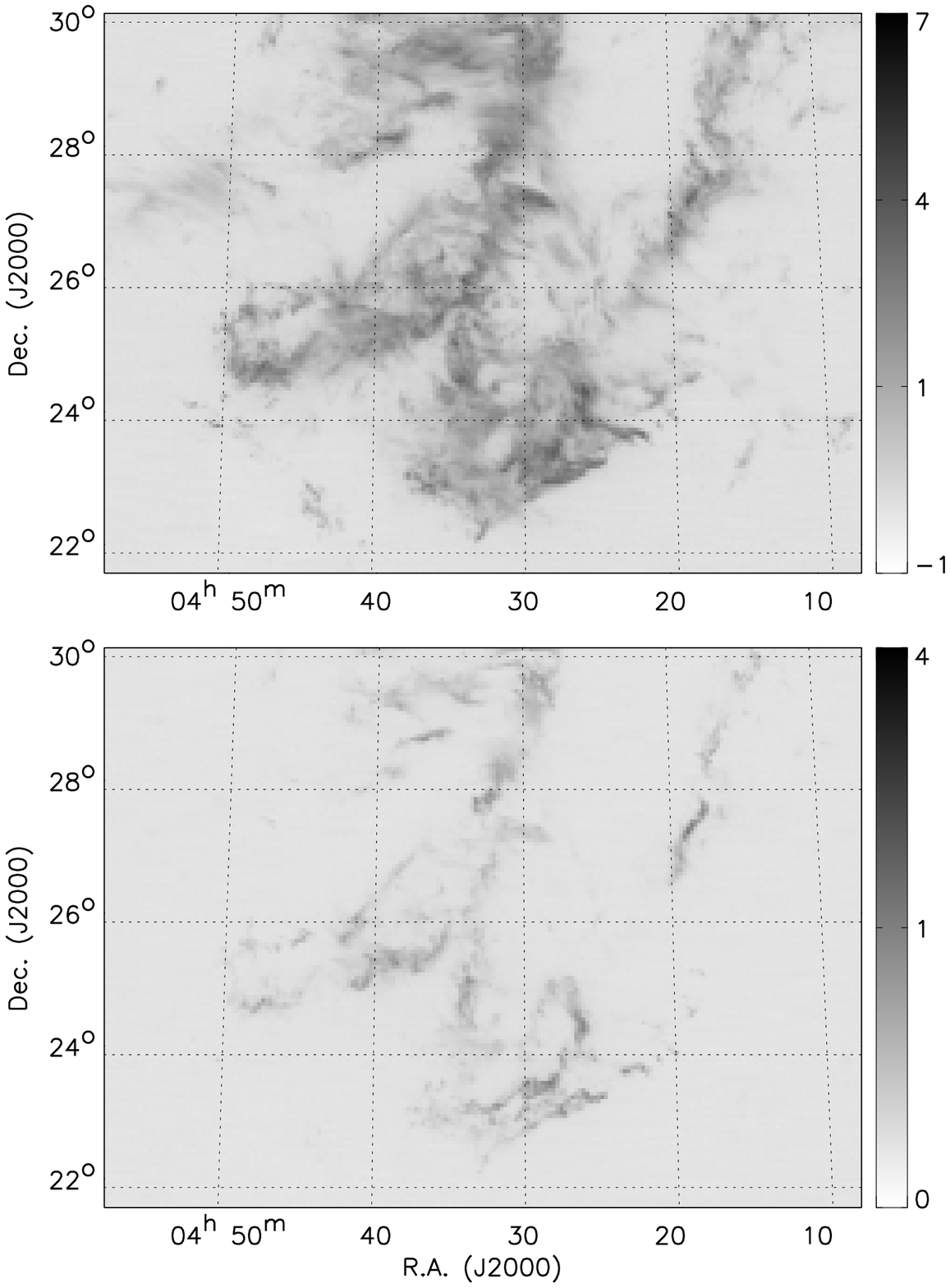}
\figcaption{Same as Figure~\ref{taurus_00} for \vlsr\ 4 to 5 \kms. \label{taurus_04}}
\end{center}
\end{figure}

\begin{figure}
\begin{center}
\epsscale{0.7}
\plotone{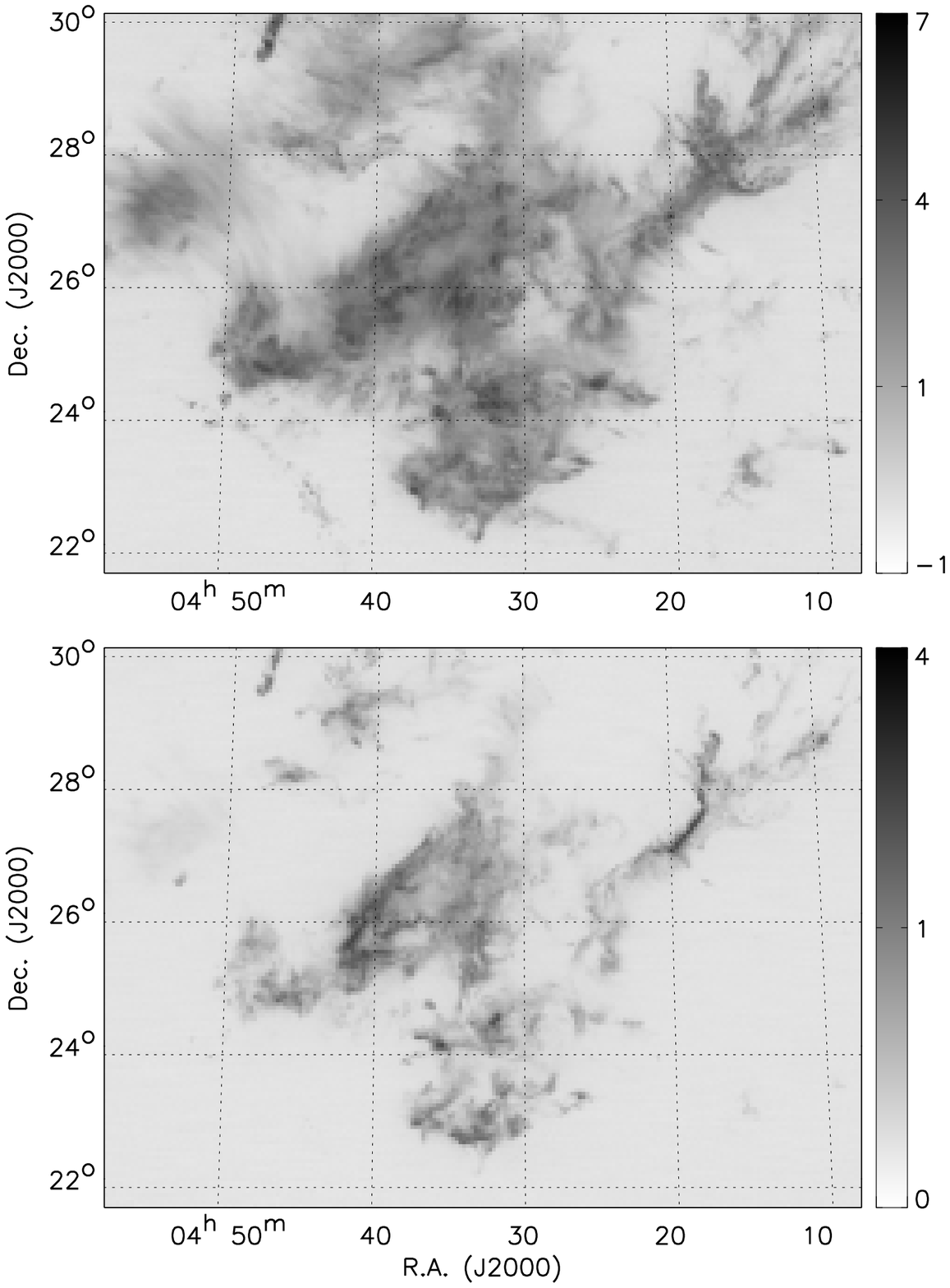}
\figcaption{Same as Figure~\ref{taurus_00} for \vlsr\ 5 to 6 \kms. \label{taurus_05}}
\end{center}
\end{figure}

\begin{figure}
\begin{center}
\epsscale{0.7}
\plotone{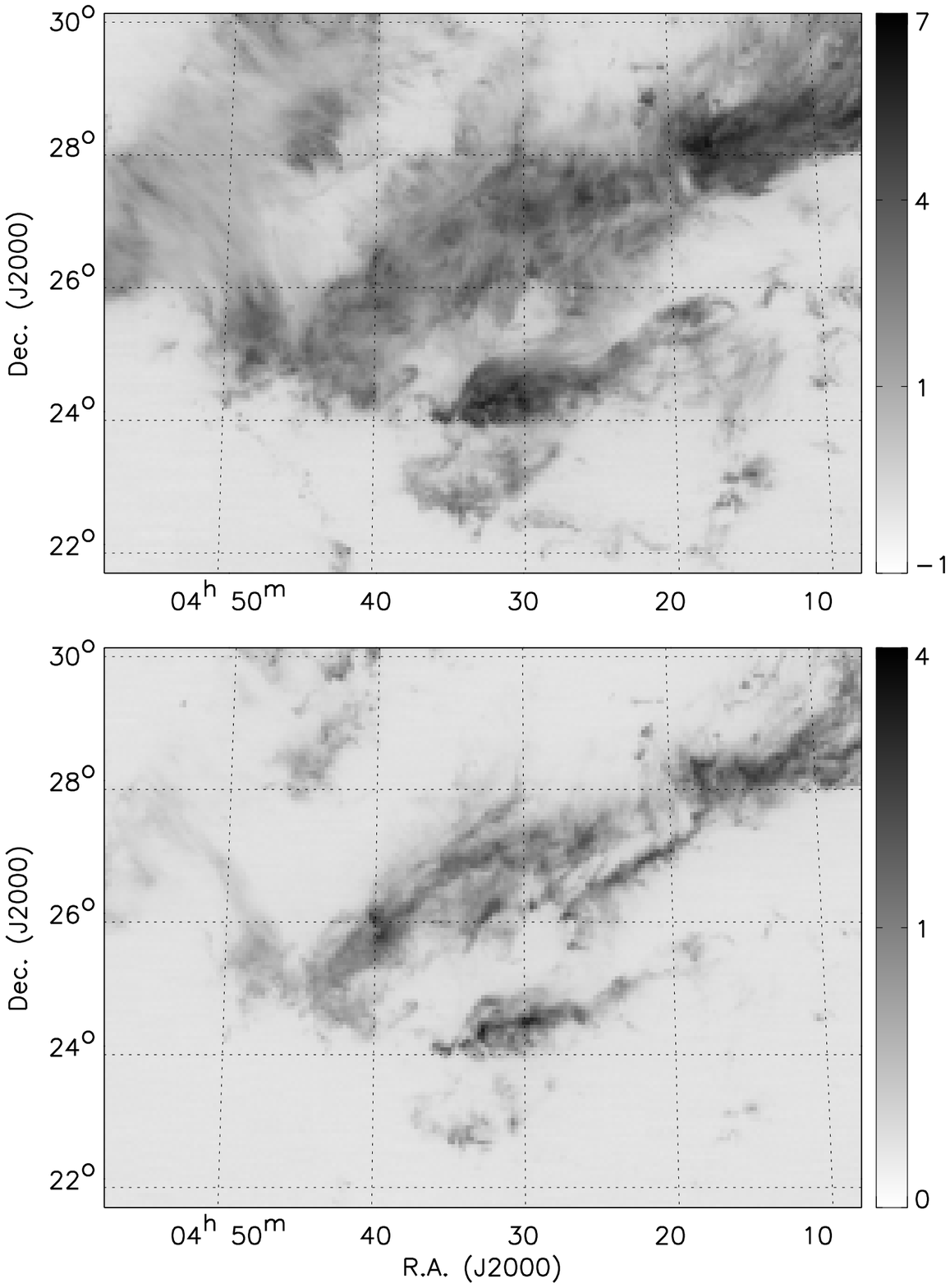}
\figcaption{Same as Figure~\ref{taurus_00} for \vlsr\ 6 to 7 \kms.\label{taurus_06}}
\end{center}
\end{figure}

\begin{figure}
\begin{center}
\epsscale{0.7}
\plotone{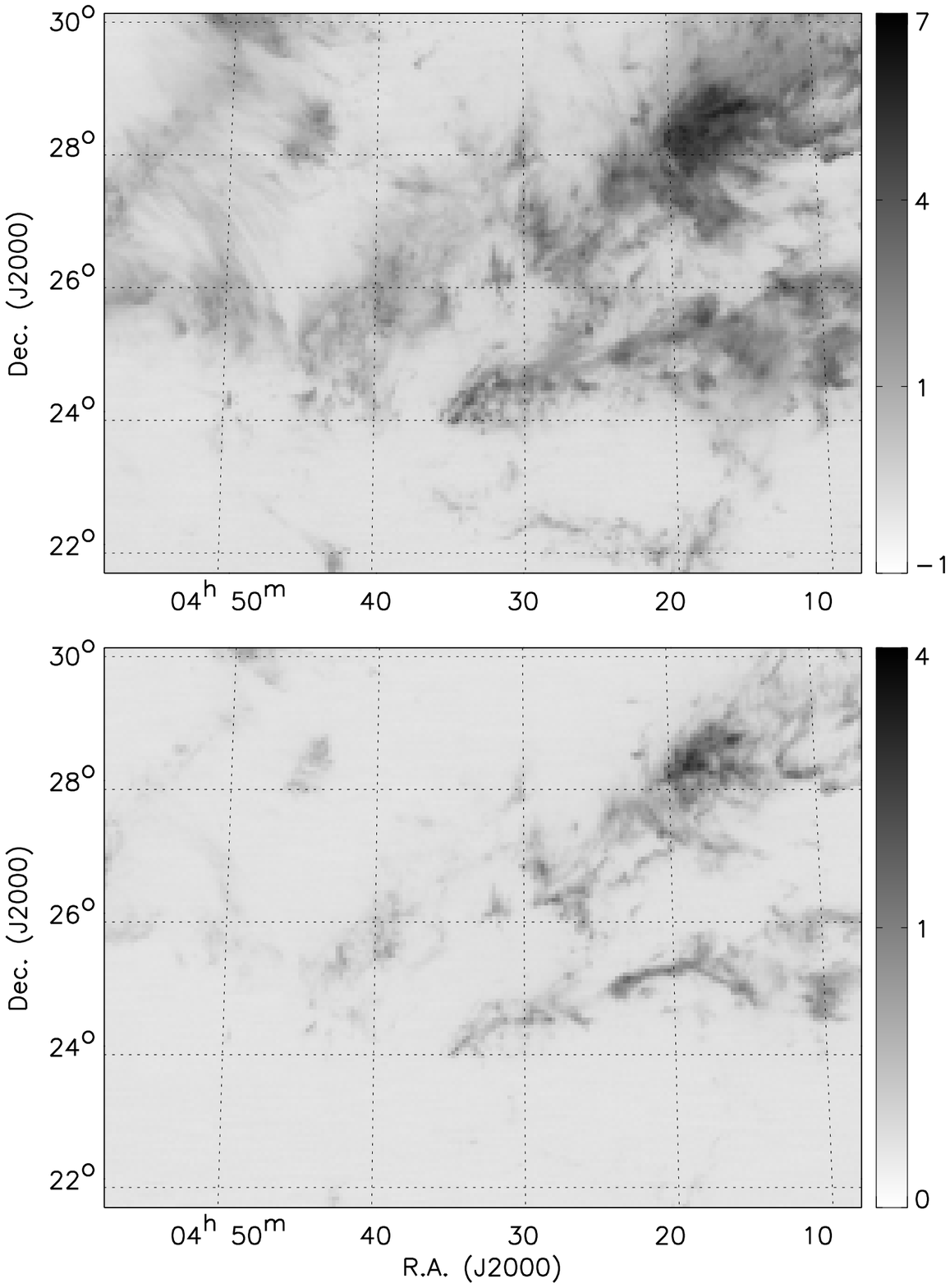}
\figcaption{Same as Figure~\ref{taurus_00} for \vlsr\ 7 to 8 \kms.\label{taurus_07}}
\end{center}
\end{figure}

\begin{figure}
\begin{center}
\epsscale{0.7}
\plotone{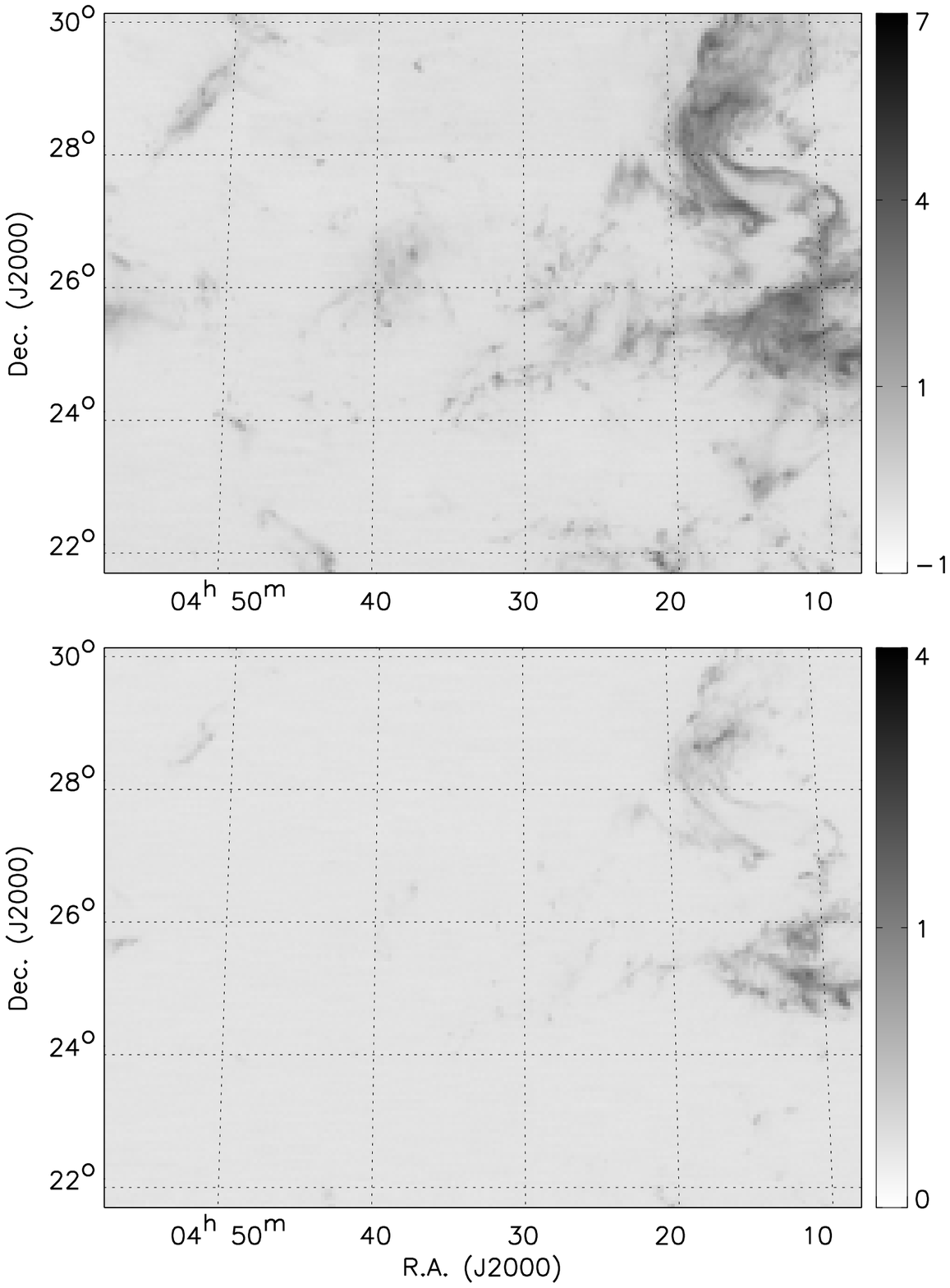}
\figcaption{Same as Figure~\ref{taurus_00} for \vlsr\ 8 to 9 \kms.
\label{taurus_08}}
\end{center}
\end{figure}

\begin{figure}
\begin{center}
\epsscale{0.7}
\plotone{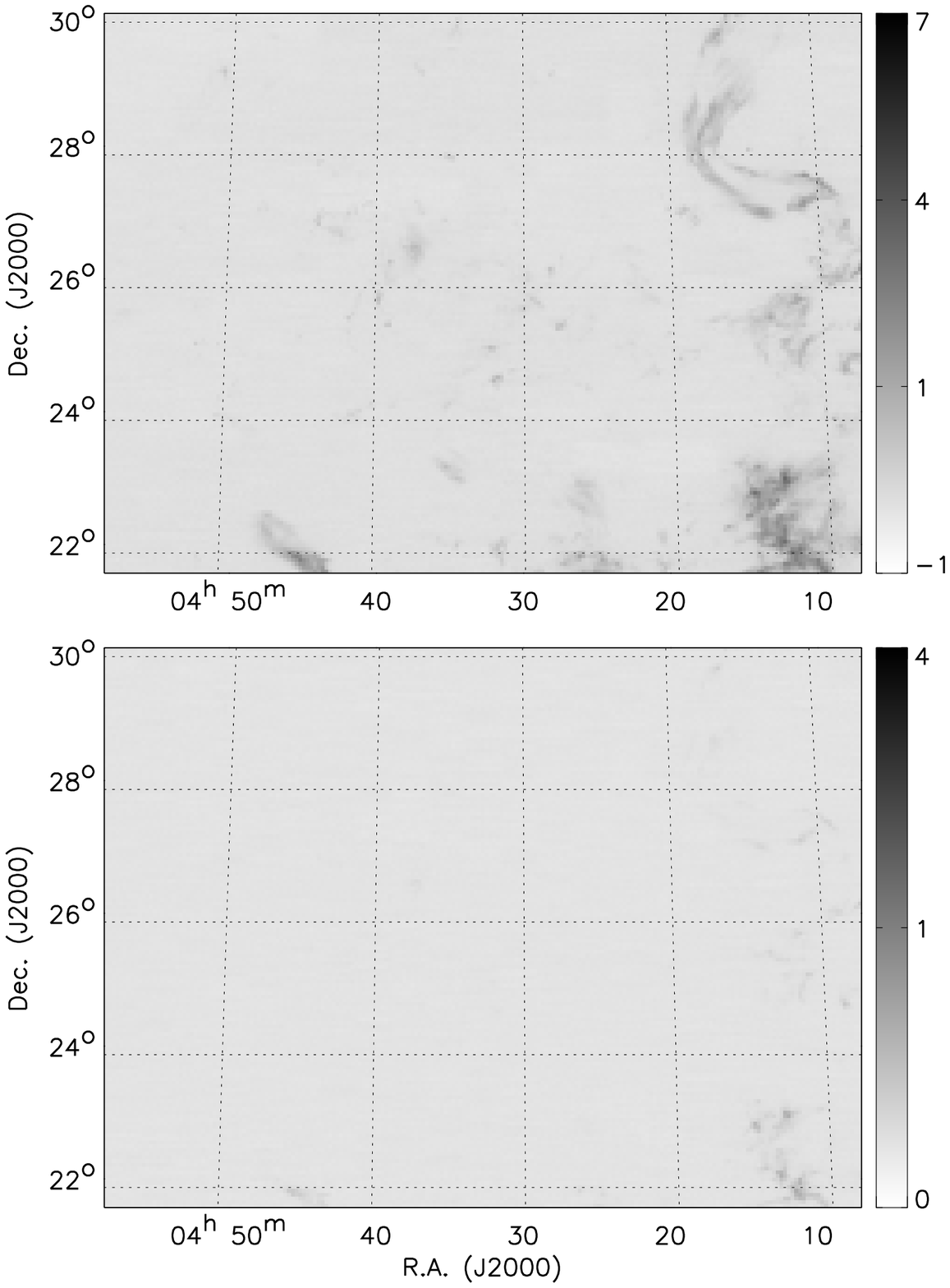}
\figcaption{Same as Figure~\ref{taurus_00} for \vlsr\ 9 to 10 \kms. \label{taurus_09}}
\end{center}
\end{figure}

\begin{figure}
\begin{center}
\epsscale{0.7}
\plotone{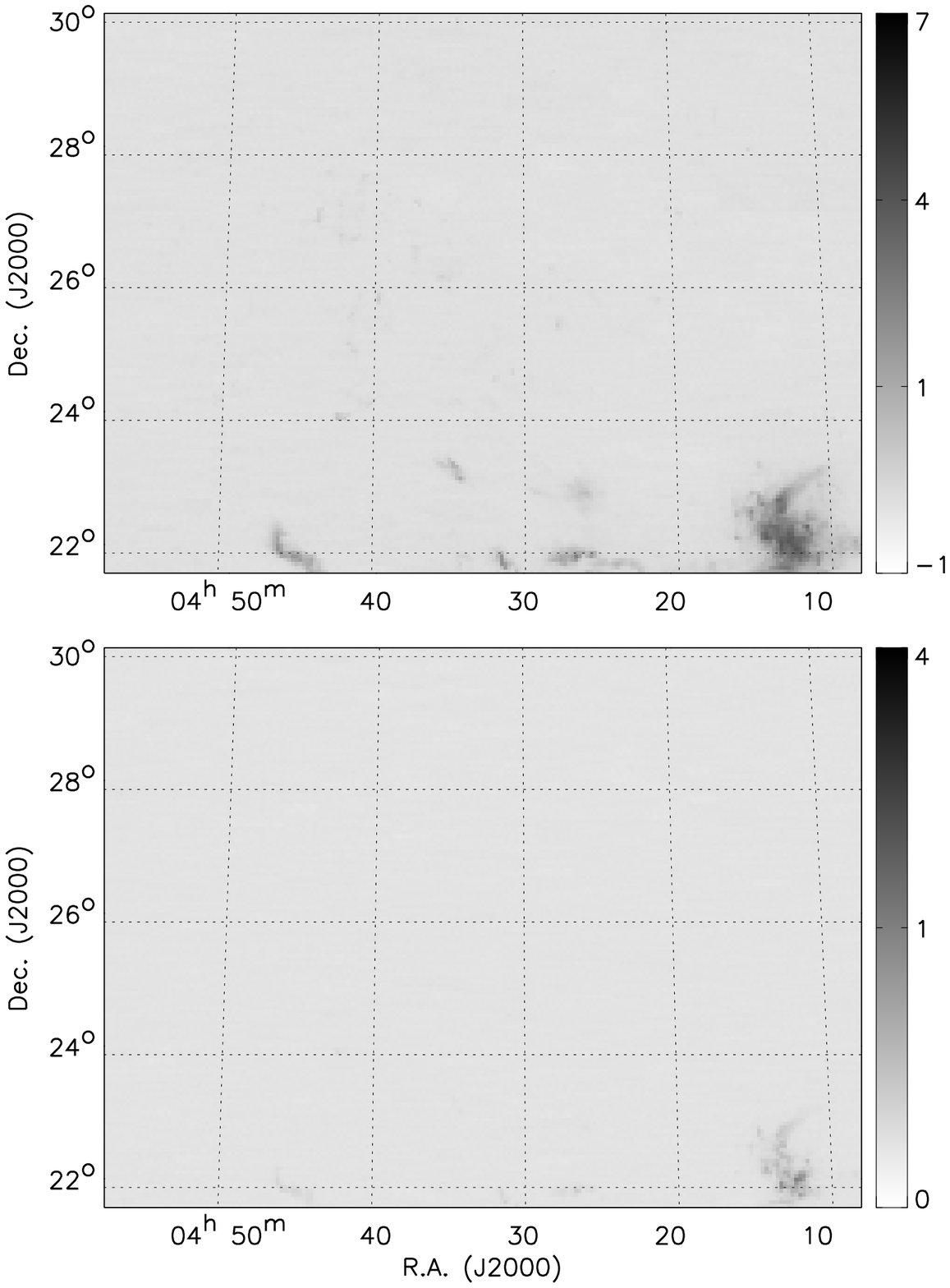}
\figcaption{Same as Figure~\ref{taurus_00} for \vlsr\ 10 to 11 \kms.
\label{taurus_10}}
\end{center}
\end{figure}

\begin{figure}
\begin{center}
\epsscale{0.7}
\plotone{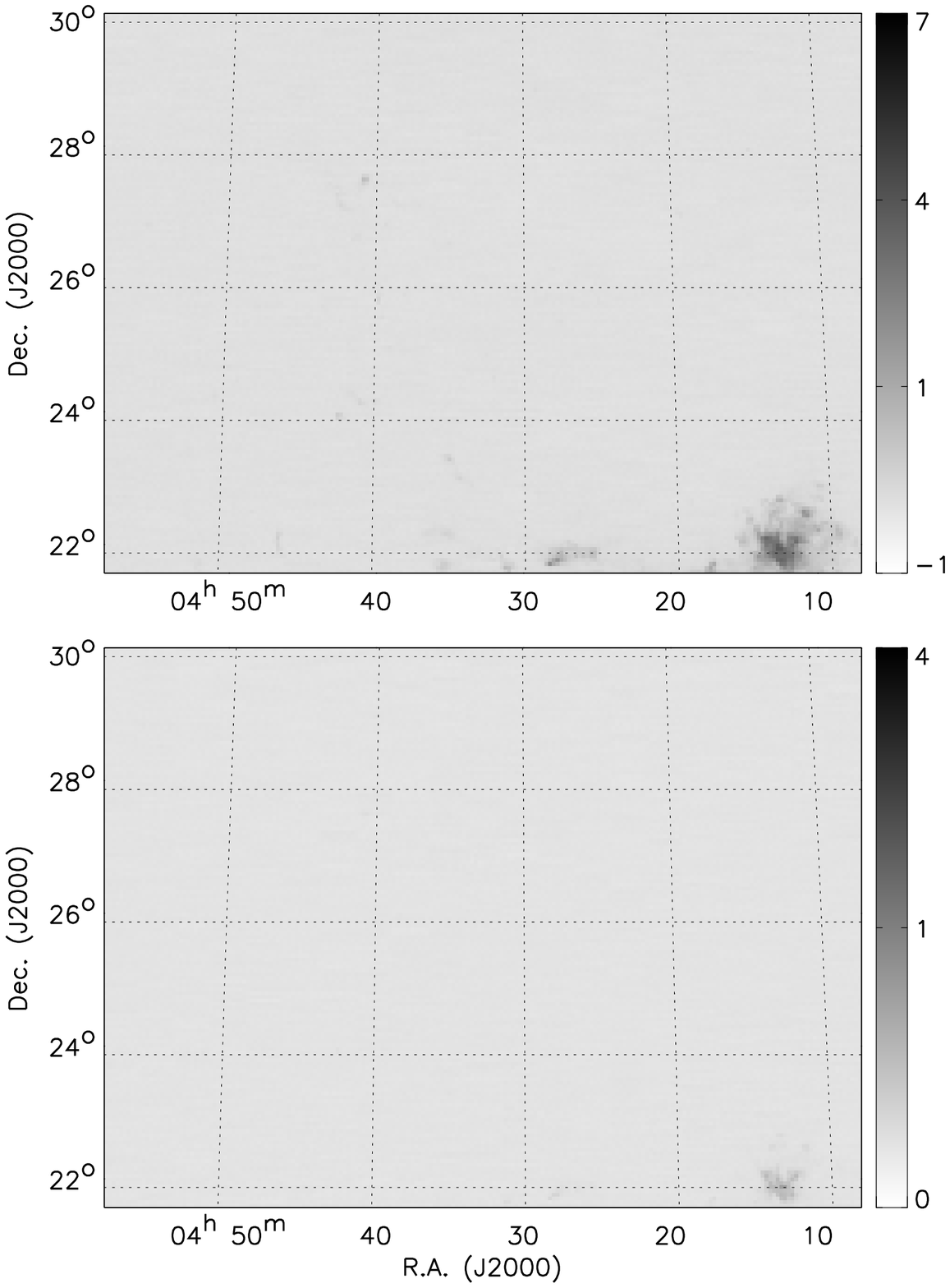}
\figcaption{Same as Figure~\ref{taurus_00} for \vlsr\ 11 to 12 \kms. \label{taurus_11}}
\end{center}
\end{figure}

\begin{figure}
\begin{center}
\epsscale{0.7}
\plotone{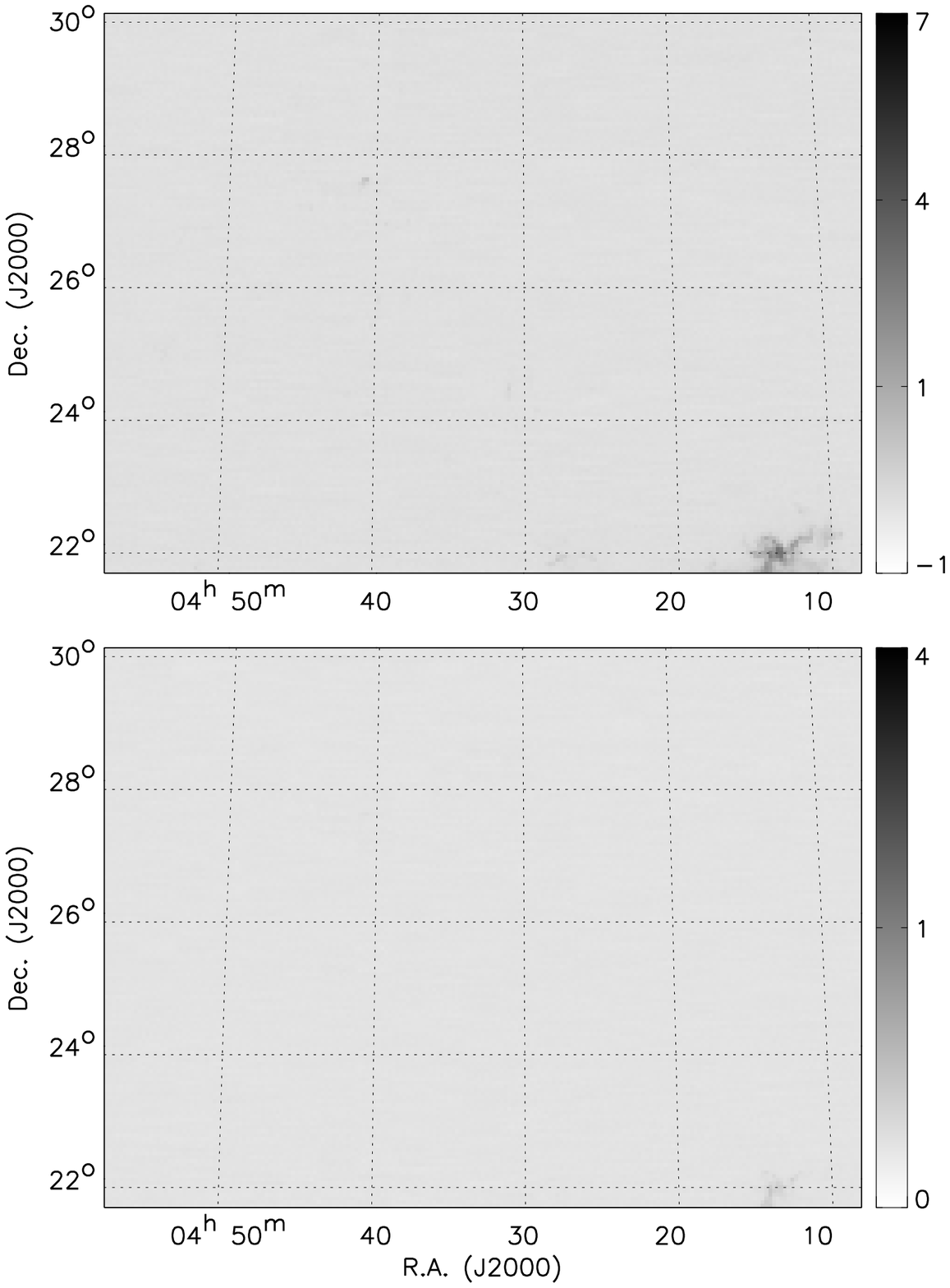}
\figcaption{Same as Figure~\ref{taurus_00} for \vlsr\ 12 to 13 \kms.
\label{taurus_12}}
\end{center}
\end{figure}

In Figures~\ref{taurus_00}-\ref{taurus_12}, we show the distribution
of \co\ and \coa\ J=1-0 emission averaged over 1 \kms\ velocity
intervals centered at \vlsr\ 0.5, 1.5, 2.5 ... 12.5 \kms.  As noted by
many previous studies \citep{brunt2004,lazarian2000}, more of the
emission is found at the higher spatial frequencies when integrating
over small velocity intervals.  The structure that is measured arises
from variations in the velocity field rather than those of density or
column density \citep{brunt2004}.  These narrow velocity integrated
images reveal stunning textural patterns and individual features that
occur over a broad range of scales.  There are regions of faint, low
surface brightness emission, most notably evident in the \co\ data.
Often, this emission component exhibits low amplitude striations that
are similar to the wind swept structures observed within terrestrial
cirrus clouds.  Owing to lower optical depth, the \coa\ emission shows
more high contrast emission originating from higher column density
regions located deeper within the cloud.  The feature located within
the southwest corner with \vlsr\ $\sim$ 10 \kms\ in Figure~16 may not
be associated with the Taurus cloud.

\section{Box Averaged Spectra}
\begin{figure}
\begin{center}
\epsscale{0.75}
\plotone{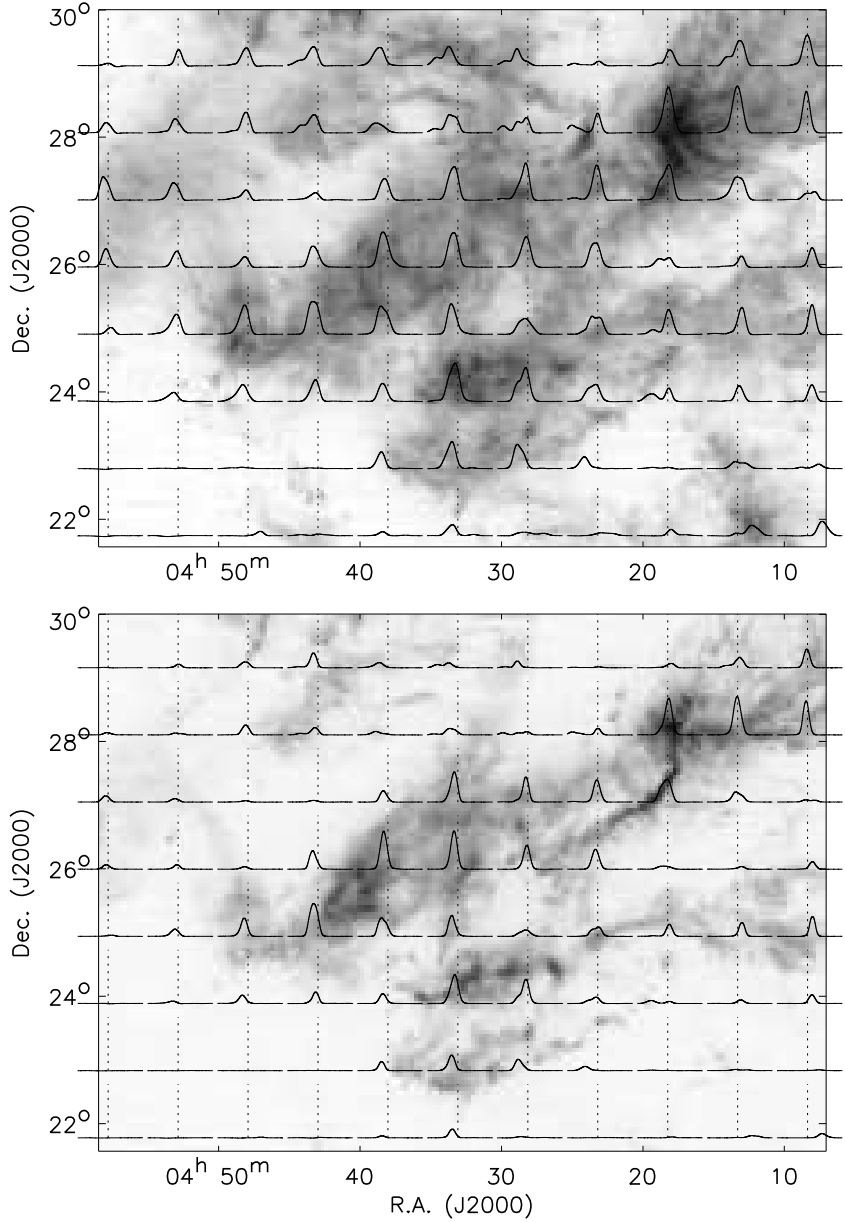}
\figcaption{Mosaic of box-averaged spectra over the 88 hard-edged
  sub-cubes for \co\ (top) and \coa\ (bottom) overlayed on images
of the integrated intensity of each isotopologue. The velocity scale on the x-axis is $0$ to $15$
  \kms, and the temperature scale ranges from
  -0.1 to 4.5 K for \co\ and -0.005 to 2.0 K for \coa. 
The vertical dotted line in each spectrum denotes \vlsr=7.0 \kms.  \label{taurus-spectra}}
\end{center}
\end{figure}

To convey the coarse velocity field of the cloud and to further
emphasize the quality of the data, we have constructed average
\co\ and \coa\ J=1-0 spectra from each 1, 1.25, or 1.56 deg$^2$ cube.
These are shown in Figure~\ref{taurus-spectra} overlayed upon images
of integrated intensity. Each spectrum is an average of 32,400,
40,500, or 50,625 individual spectra with typical rms values of
0.014 K for \co\ and 0.006 K for \coa\ (T$_A^*$).  These rms values
are greater than one would expect averaging this number of spectra
each with the respective median values shown in Figure~\ref{hist_rms}.
This discrepancy arises from spatially correlated noise imposed on
spectra sharing a common reference measurement in the OTF data
collection scheme.  The average spectra demonstrate excellent baseline
fidelity with no evidence for any systematic noise contribution.

The average spectra do reveal several significant properties of the Taurus
velocity field.  The previously established large scale velocity
gradient across the Taurus Molecular Cloud is apparent in the systemic
shift of the spectra from blueshifted velocities on the eastern side
to redshifted velocities on the western portion of the cloud.  In
addition, many of the average spectra exhibit asymmetries and multiple
velocity components that attest to the complex structure along the line of sight.

\section{Emission Statistics}

The \co\ and \coa\ data cubes of Taurus provide panoramic views of the
structure of a 10$^4$ \msun\ molecular cloud.  The resident
information is sufficiently vast to require statistical descriptions
of the data that may offer insight to the prevailing conditions or
state of the cloud.  
The cumulative probability
density function is defined to be
$$ P(f_\circ) = { {\int \int dx dy F(x,y,f_\circ) } \over { \int \int dx dy F(x,y,f_{max})} } \eqno (2), $$
where 
\[ F(x,y,f_\circ) = \left\{  
\begin{array}{ll} 
0 & \mbox{for $ f(x,y) > f_\circ$} \\
1 & \mbox{for $f(x,y) \le f_\circ$},\\
\end{array} 
\right. \]\\ with $f(x,y)$ the 2 dimensional distribution of some
measured parameter, $f_\circ$ a moving threshold of that parameter,
and $f_{max}$ the maximum value of the measured parameter in the whole
map.  The weighted cumulative PDF biases each bin of $f(x,y)$ by the
total signal, $\int T(x,y,v)dv$, of all contributing pixels,
$$ P_w(f_\circ) = { {\int \int dx dy F(x,y,f_\circ) \int T(x,y,v)dv } \over { \int \int dx dy \int T(x,y,v) dv } } \eqno(3) $$

\begin{figure}
\begin{center}
\plotone{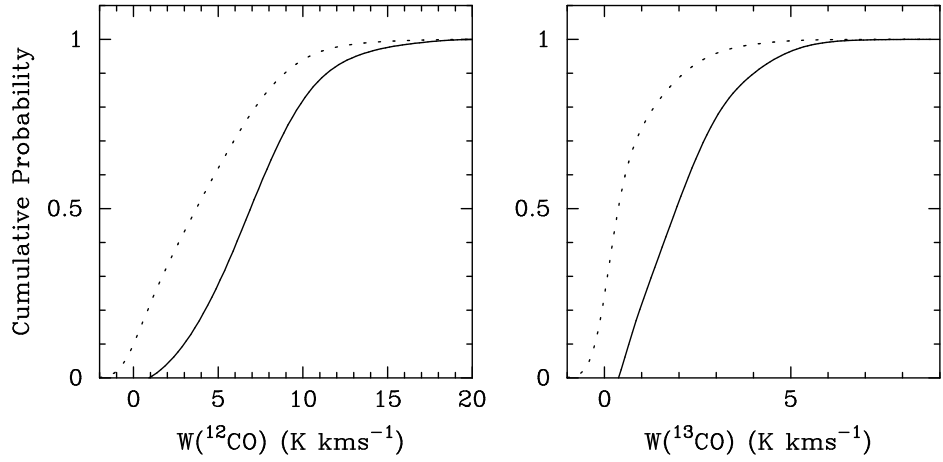}
\figcaption{Weighted (solid line) and unweighted (dotted line) cumulative
  probability density functions as a function of \co\ and
  \coa\ integrated intensity (denoted as W(\co) and W(\coa)
  respectively).  The unweighted and weighted cumulative PDFs describe
  the fractional contribution by projected area and integrated
  intensity emission respectively.
\label{hist_wttint}}
\end{center}
\end{figure}

\begin{figure}
\begin{center}
\plotone{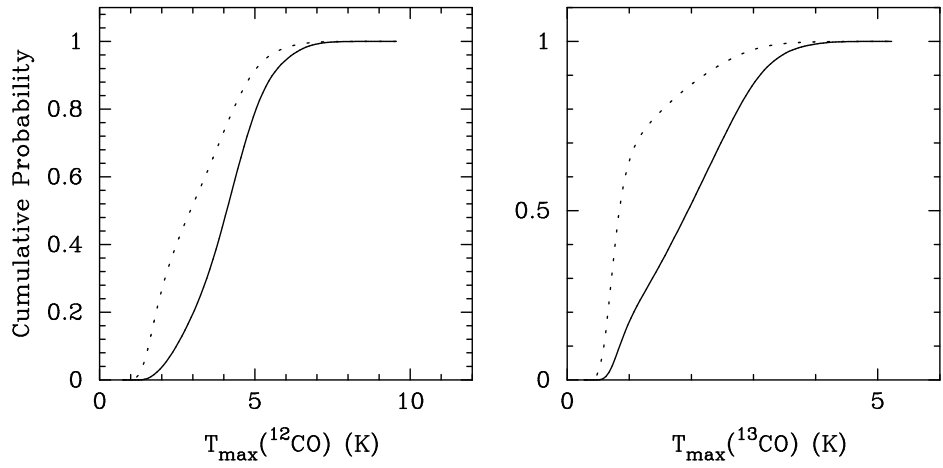}
\figcaption{Weighted (solid line) and unweighted (dotted line) cumulative
  probability density functions as a function of \co\ and \coa\ peak
  intensity.  
\label{hist_wttmax}}
\end{center}
\end{figure}

Figures~\ref{hist_wttint} and \ref{hist_wttmax} show the cumulative
PDF and weighted cumulative PDF for integrated intensity (denoted as
W(\co) and W(\coa) respectively), and peak temperature distributions
(denoted as T$_{max}$(\co) and T$_{max}$(\coa) respectively). The
weighting function used is the corresponding integrated intensity.
The unweighted cumulative PDF shows the cumulative fraction of
projected area as a function of integrated intensity
(Figure~\ref{hist_wttint}) and peak temperature
(Figure~\ref{hist_wttmax}). The weighted cumulative PDF shows the
cumulative fraction of integrated intensity as a function of
integrated intensity (Figure~\ref{hist_wttint}) and peak temperature
(Figure~\ref{hist_wttmax}).  The median values for the weighted PDFs
of integrated intensity are 6.9 and 1.9 \kkms\ for \co\ and
\coa\ respectively.  Similarly, for peak temperature, the weighted
median values are 4.1 K and 1.9 K. These figures demonstrate that much
of the \co\ and \coa\ signal originates from lines of sight within the
low surface brightness regime of the Taurus cloud.  For example, the
weighted cumulative PDFs demonstrate that half of the \co\ flux of a
cloud is emitted within the low surface brightness portion of the
Taurus molecular cloud having \co\ integrated intensity less than 6.9
\kkms\ and peak temperature $\leq 4.1$~K. In terms of projected area,
only $\sim 20$\% of the cloud's area has integrated intensity $\geq
6.9$ \kkms\ or peak temperature $\geq 4.1$~K in \co.


\section{Conclusions}
The FCRAO Survey of the Taurus Molecular Cloud is a powerful set of
data to investigate interstellar gas dynamics and the star formation
process with high spatial resolution and spatial dynamic range.  It
offers a valuable complement to observations at other wavelengths that
probe the dust component and the population of young stellar objects.
In this paper, we summarized the instrumentation, data collection and
processing procedures used in the survey. We also characterized the
noise and signal distributions of the survey.  The overall structure
of the cloud, its column density distribution and mass, and
relationship with the magnetic field are discussed in
\cite{goldsmith2008}.

\acknowledgements This work was supported by NSF grant AST 05-40852 to
the Five College Radio Astronomy Observatory, NSF grant AST-0407019 to
Cornell University, and by the Jet Propulsion Laboratory, California
Institute of Technology.  We thank Yvonne Tang and Marko Krco for for
assistance with observations.

\clearpage

\clearpage

\clearpage

\clearpage

\clearpage

\clearpage

\clearpage

\clearpage

\clearpage

\clearpage

\clearpage

\clearpage

\clearpage

\clearpage

\clearpage

\clearpage

\clearpage

\clearpage

\clearpage

\clearpage

\clearpage

\clearpage

\end{document}